\documentclass[a4paper, oneside, twocolumn, notitlepage, 10pt]{extarticle_ecoc}
\usepackage{ecoc}
\usepackage{subfig}

\def\fw{f_{\mathrm{w}}}

\def\e{{\mathrm{e}}}
\def\j{{\mathrm{j}}}
\def\Np{N_{\mathrm{p}}}

\DeclareMathOperator*{\argmax}{argmax}
\newcommand{\appropto}{\mathrel{\vcenter{
  \offinterlineskip\halign{\hfil$##$\cr
    \propto\cr\noalign{\kern2pt}\sim\cr\noalign{\kern-2pt}}}}}
    
\makeatletter
\def\blfootnote{\gdef\@thefnmark{}\@footnotetext}
\makeatother
\begin{document}
\selectlanguage{english}    


\title{Bayesian Phase Search for Probabilistic Amplitude Shaping}%


\author{
    Mohammad Taha Askari and Lutz Lampe
}

\maketitle                  


\begin{strip}
 \begin{author_descr}

   Department of Electrical and Computer Engineering, University of British Columbia, Vancouver, BC V6T 1Z4, Canada,
   \textcolor{blue}{\uline{mohammadtaha@ece.ubc.ca}}

 \end{author_descr}
\end{strip}

\setstretch{1.1}
\renewcommand\footnotemark{}
\renewcommand\footnoterule{}


\begin{strip}
  \begin{ecoc_abstract}
    We introduce a  Bayesian carrier phase recovery (CPR) algorithm which is robust against low signal-to-noise ratio scenarios. It is therefore effective for phase recovery for probabilistic amplitude shaping (PAS). Results validate that the new algorithm overcomes the degradation experienced by blind phase-search CPR for PAS. \textcopyright2023 The Author(s)
  \end{ecoc_abstract}
\end{strip}


\section{Introduction}

\blfootnote{This work was supported by Huawei Tech., Canada, and enabled in part through support provided by the Digital Research Alliance of Canada (www.alliancecan.ca).}

Probabilistic amplitude shaping (PAS) \cite{bocherer2015bandwidth} has been widely adopted as a mechanism to increase data rate and range in optical fiber communication systems. In addition to offering the usual linear shaping gain, short blocklength PAS also decreases the nonlinear signal-to-signal interactions along the optical fiber resulting in a nonlinear shaping gain \cite{fehenberger2019analysis, askari2023probabilistic, civelli2023nonlinear}.
However, while PAS improves the nonlinear tolerance of the optical fiber communication system, it has the potential to compromise the performance of other  digital signal processing modules used in the transmission system. In particular, the interplay of carrier phase recovery (CPR) using the blind phase search (BPS) algorithm and PAS has been shown to be problematic \cite{mello2018interplay}.

 BPS is  a classical CPR method widely employed for its hardware-efficient implementation \cite{pfau2009hardware}. It is based on minimizing the total Euclidean distance between the rotated symbols and the nearest constellation points for a window of consecutive symbols. Rotational ambiguity in quadrature amplitude modulation (QAM) constellations can be resolved through differential coding \cite{weber1978differential} or pilot symbols  \cite{magarini2012pilot}. As a decision-directed CPR method, BPS does not perform well in the low signal-to-noise ratio (SNR) regime \cite{yankov2015low}. For such scenarios,  the work in \cite{yankov2015low} introduced a phase-noise tracking algorithm, which is based on the assumption of a mixture of von Mises distributions for the phase at each time step, and message passing for deriving the posterior probability of input symbols.

The BPS performance degrades substantially when combined with PAS, especially in the near-optimal region for the PAS shaping parameter \cite{mello2018interplay}. This is due to the higher probability of low-energy QAM symbols, which have a larger phase gap to the nearest adjacent constellation point. Therefore, erroneous decisions used in the decision-directed BPS are more detrimental to phase tracking. Although increasing the BPS window size partially alleviates this effect, excessively large window sizes render phase-noise tracking too slow and make BPS computationally complex \cite{mello2018interplay}. The work in \cite{di2020low} suggested a two-stage CPR algorithm specially designed for PAS. It incorporates the symbols' prior distribution in the decision process and outperforms BPS, but only for constellations with relatively low entropy values. Furthermore, it has recently been shown that BPS provides a  nonlinear gain similar to that of short-blocklength PAS in the high SNR regime, but it fails to do so for low SNR due to the required large window size \cite{civelli2023nonlinear}. 

In this paper, we suggest a new CPR algorithm that is effective for PAS transmission. It combines the relative simplicity of BPS with an additional phase prior that is obtained from few, periodically introduced pilot symbols. Such pilot symbols are already part of the transmission system as they are  used for adaptive equalization at the receiver. Likewise, BPS requires pilots to resolve phase ambiguities. Since the new method employs a phase prior as well as a symbol prior accounting for PAS, we refer to it as Bayesian phase search (BaPS). Our numerical results show that BaPS works well for PAS even with short window sizes and outperforms BPS. 
 Furthermore, BaPS automatically prevents cycle slips.
\section{System Model}
For the derivation of BaPS, we consider the simplified phase noise model for a CPR as \cite{yankov2015low} 
\vspace*{-1mm}
\begin{equation}
    y_k = x_k \e^{\j \theta_{k}} + n_{k},
    \label{eq:obv_model}
\end{equation}
where $x_k \in {\cal X}$ and $y_k$ denote the transmitted symbol from the constellation $\cal X$ and received sample at the discrete time $k$, respectively. $n_k$ is the complex additive noise at each time step, which is independent of $x_k$ and $\theta_k$ and follows a complex Gaussian distribution with zero mean and  variance $\sigma^2$, i.e.,\ $n_{k} \sim \mathcal{N}(0,\sigma^2)$. We assume a Wiener process for the phase noise as 
\vspace*{-1mm}
\begin{equation}
    \theta_k = \theta_{k-1} + v_k,
    \label{eq:wiener}
\end{equation}
where $v_{k} \sim \mathcal{N}(0,\Delta^2)$, and $\Delta^2$ is the total variance of phase noise due to fiber nonlinearity and the laser phase noise at the transmitter and receiver. 
\section{Bayesian Phase Search}
We start with the simplifying assumption that the phase noise $\theta_k$ is approximately constant for several symbols around the symbol of interest, i.e.,\ $y_{k-N}^{k+N} = \{ y_{k-N}, \dots, y_k, \dots, y_{k+N} \}$, 
where
$N$ is the \textsl{one-sided} window size for the phase search. Thus, the maximum a-posteriori probability (MAP) estimation of phase $\theta_k$ is derived as
\begin{equation}
\hat{\theta}_k  = \argmax_{\theta_k} \ p(\theta_k | y_{k-N}^{k+N}).
\end{equation}
\vspace*{-1mm}
The posterior distribution is proportional to
\vspace*{-1mm}
\begin{equation}
    p(\theta_k | y_{k-N}^{k+N}) 
               \propto p(\theta_k) \prod_{n=k-N}^{k+N} p(y_n|\theta_k), \label{eq:independent}
\end{equation}
which follows from Bayes rule 
and conditional independence of $y_n$ given $\theta_k$. We use the law of total probability and the max-sum approximation to simplify the conditional probability $p(y_n|\theta_k)$ 
as
\begin{equation}
           p(y_n|\theta_k) =\! \sum_{x_n \in {\cal X}} p(y_n|x_n, \theta_k) p(x_n) 
           \appropto 
          p(y_n|\hat{x}_n, \theta_k), \label{eq:decision}
\end{equation}
\vspace*{-1mm}
where 
\vspace*{-1mm}
\begin{equation}
    \hat{x}_{n,k} = \argmax_{x_n \in {\cal X}} \frac{-|y_n \e^{-\j\theta_k} - x_n|^2}{\sigma^2} + \log p(x_n).
    \label{eq:paps_decision}
\end{equation}
Finally, plugging $p(y_n | \theta_k)$ from \eqref{eq:decision} into \eqref{eq:independent}, 
the BaPS phase noise estimate at time step $k$ is
\vspace*{-1mm}
\begin{equation}
\hat{\theta}_k  = \argmax_{\theta_k} \ \log p(\theta_k) - \!\!\sum_{n=k-N}^{k+N} \!\!\frac{|y_n \e^{-\j\theta_k} -{\hat x_{n,k}}|^2}{\sigma^2}.
\label{eq:paps}
\end{equation}

We note that similar to BPS, the maximization in \eqref{eq:paps} can be carried out using a set of test phases. Hence, BaPS can benefit from efficient implementations of BPS. Different from BPS, BaPS (i) considers test phases in the complete unit circle $[ -\pi, \pi )$, (ii) incorporates the symbols prior probability $p(x_n)$ in the decision \eqref{eq:paps_decision}, and (iii) considers the phase prior probability $p(\theta_k)$ in the phase recovery \eqref{eq:paps}. As typical for Bayesian estimation, for a small number of observations the prior term  in \eqref{eq:paps} matters. Thus, we expect that BaPS will outperform BPS  for small window sizes $N$. The second term in \eqref{eq:paps} will dominate for larger window sizes $N$, and thus we expect the BPS and BaPS performances to become similar. 

\emph{Prior distribution:} 
As the phase noise in \eqref{eq:wiener} follows a wrapped Gaussian distribution, we approximate it by the von Mises distribution\footnote{$I_0$ is the zeroth-order modified Bessel function.}
\vspace*{-1mm}
\begin{equation}
    p(\theta_k) = \frac{\exp(\mathrm{Re}[z_k \exp(-\j\theta_k)])}{2\pi I_0(|z_k|)},
    \label{eq:von}
\end{equation}
where $z_k$ is a complex parameter of the distribution that is updated as\cite{barbieri2007soft}
\vspace*{-1mm}
\begin{equation}
    z_{k+1} = \frac{z_k}{1+\Delta^2|z_k|}.
    \label{eq:param_upd}
\end{equation}

In order to initialize the parameter in \eqref{eq:param_upd}, and to estimate the variances $\Delta^2$ and $\sigma^2$, we perform maximum likelihood phase estimation 
\cite{zhou2010improved} 
\begin{equation}
    \vspace*{-1mm}
    \!\!{\hat \theta}_p =  \angle \!\!\!\! \sum_{n=\max ( 1, p-\frac{L}{2} )}^{\min (\Np, p+\frac{L}{2})} \!\!y_n x_n^*, \quad p \in \{ 1, \dots, \Np \},
    \label{eq:ml_pilot}
\end{equation}
using $\Np$ pilot symbols and estimation window size $L$. In \eqref{eq:ml_pilot}, $(.)^*$ and $\angle(.)$ denote the complex conjugate and angle of a complex number, respectively. 
The parameter estimates follow as
\vspace*{-1mm}
\begin{subequations}
    \begin{align}
       \hat{\sigma}^2 &= 
        \frac{1}{\Np} \sum_{p=1}^{\Np} |y_p - x_p \e^{\j{\hat \theta}_p}|^2,  \label{eq:est_a} \\
        \hat{\Delta}^2 &= 
            \frac{1}{\Np-1} \sum_{p=2}^{\Np} ({\hat \theta}_p - {\hat \theta}_{p-1})^2, \label{eq:est_b}  \\
        z_{\Np+1} &= \frac{1}{{\hat \Delta}^2} \e^{\j {\hat \theta}_{\Np}}. \label{eq:est_c}
    \end{align}
    \label{eq:est}
    \vspace*{-5mm}
\end{subequations}

\section{Numerical Results}
We compare the performances of BaPS and BPS for the additive white Gaussian noise (AWGN) channel scenario from \cite{mello2018interplay} and the optical fiber communication scenario  from \cite{askari2023probabilistic}.


\begin{figure*}[t]
\centering
\subfloat{\includegraphics[width=0.47\textwidth]{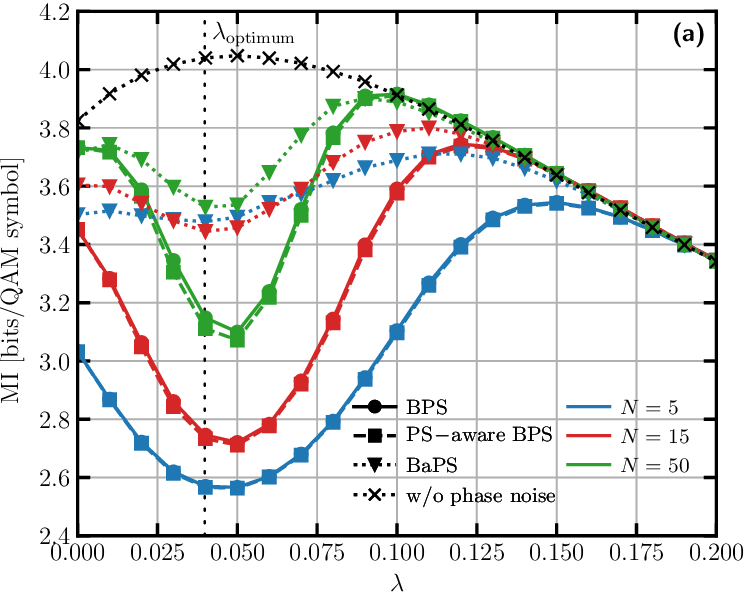}
\label{fig:awgn_a}}
\hfil
\subfloat{\includegraphics[width=0.47\textwidth]{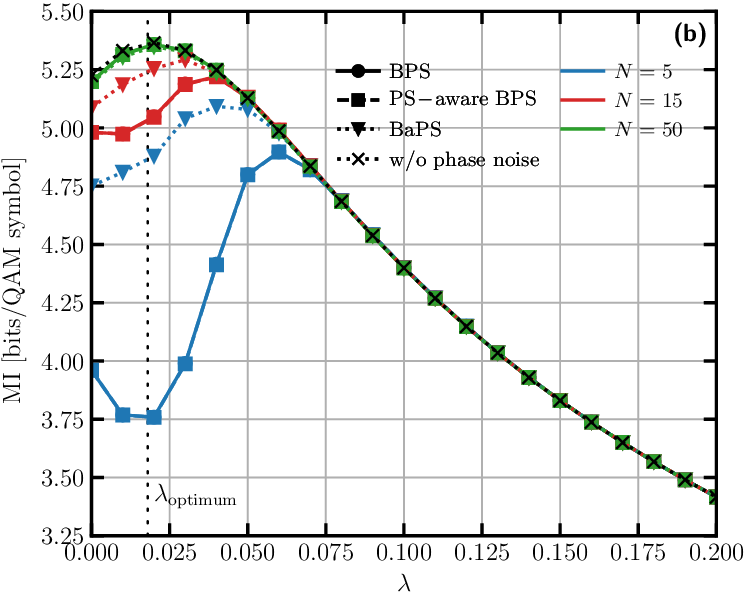}}%
\label{fig:awgn_b}
\caption{MI vs. MB parameter at SNR equal (a) 12 dB and (b) 17 dB. Wiener phase noise with $\fw=200$~kHz at $50$~GBd. The case of transmission without phase noise is shown as a reference.}
\label{fig:awgn}
\end{figure*}

\emph{AWGN channel:} We generate $2^{18}$ symbols with $50$~GBd baudrate using  a $64$-QAM PAS system with an ideal distribution matcher for a Maxwell Boltzmann (MB) distribution with parameter $\lambda$.  The symbols are transmitted through an AWGN channel with a Wiener phase noise with $
\fw=200$~kHz, and at the receiver  CPR with BaPS and BPS are carried out. 
 As BPS searches over a set of test phases in the range $[0,\pi/2)$ and BaPS searches over the range $[-\pi,\pi)$,  we used $60$ test phases for BPS and $240$ test phases for BaPS to have a fair comparison. 
Moreover, to resolve the phase ambiguity after BPS, we apply the supervised cycle slip correction \cite{civelli2023nonlinear, mello2018interplay}.  This is not needed for BaPS. $\Np=50$ pilots are interleaved every $2,000$  transmitted symbols used in \eqref{eq:ml_pilot} and \eqref{eq:est}. 
The parameter $L$ is optimized for the best performance at each CPR window size $N$.

Fig.~\ref{fig:awgn} shows the 
mutual information (MI), computed based on the method in \cite{renner2017experimental}, versus the MB parameter $\lambda$ for SNRs of (a) $12$~dB and (b) $17$~dB. As BPS was originally introduced for uniform QAM constellations, we also consider a probabilistic shaping aware (PS-aware) version of BPS, in which the prior probabilities are incorporated in the decision process using \eqref{eq:paps_decision}. Fig.~\ref{fig:awgn} shows that both versions of BPS fail to recover the carrier phase at $\lambda$-values around the optimal shaping parameter. While this effect can be mitigated by increasing $N$, this in turn affects the reliability of the CPR algorithm for fast phase drift. BaPS is consistently superior to BPS, especially for low SNR and short window sizes. 


\begin{figure}[t]
   \centering
    \includegraphics[width=\linewidth]{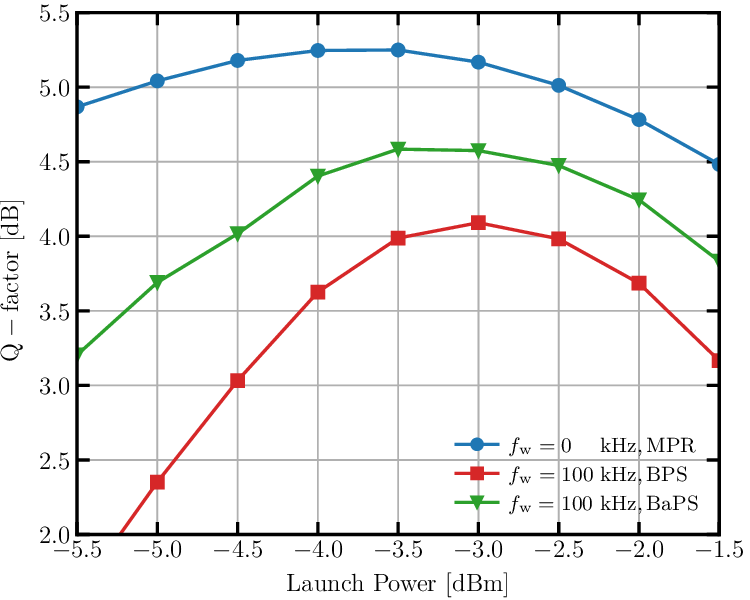}
    \caption{Q-factor vs.\ launch power. B(a)PS with  $N=50$.}
    \label{fig:fiber}
\end{figure}

\emph{Optical fiber channel:} 
We simulate a single-polarization wavelength division multiplexing system with $11$ channels, $32$~GBd baudrate, and $50$~GHz channel spacing. $256$-QAM symbols are generated using PAS with an ideal MB distribution matcher with a rate of $2.4$~bits/amplitude ($\lambda\!\!=\!0.019$). Laser phase noise with $\fw\!=\!100$~kHz is applied at the transmitter and receiver. $15$~spans of $80$~km of  a standard single-mode fiber with chromatic dispersion (CD) parameter $17$~ps/nm/km,  nonlinearity parameter $1.37$~W$^{-1}$km$^{-1}$, and loss $0.2$~dB/km are used. Erbium-doped fiber amplifier with a $6$~dB  noise figure is deployed at the end of each span. At the receiver, electronic CD compensation is followed by CPR, and then the performance for the central channel is computed. 

Fig.~\ref{fig:fiber} compares the Q-factor performances of BPS and BaPS with window size $N\!\!=\!50$ and equal test phase resolution $\frac{\pi}{120}$. As a reference, we also include the case of ideal transmission without laser phase noise and mean phase rotation (MPR) compensation. We observe that BaPs outperform BPS in terms of Q-factor for all launch powers. This is consistent with the results for the AWGN channel in Fig.~\ref{fig:awgn} and confirms the effectiveness of BaPS when combined with PAS. Moreover, the optimal launch powers are higher than for the MPR curve, which is due to the compensation of nonlinear phase noise by B(a)PS.

\section{Conclusions}
We derived BaPS algorithm for phase noise tracking based on the MAP phase estimation. 
BaPS enjoys the same hardware-efficient implementation as BPS, while it also provides advantages over BPS for shorter estimation windows and low SNRs, which makes it a promising solution for PAS and strong laser phase noise scenarios.


\printbibliography

\vspace{-4mm}

\end{document}